# Momentum matching and band-alignment type in van der Waals heterostructures: Interfacial effects and materials screening

Yue-Jiao Zhang, Yin-Ti Ren, Xiao-Huan Lv, Xiao-Lin Zhao, Rui Yang, Nie-Wei Wang, Chen-Dong Jin, Hu Zhang, Ru-Qian Lian, Peng-Lai Gong, Rui-Ning Wang, Jiang-Long Wang*, and Xing-Qiang Shi*

Key Laboratory of Optic-Electronic Information and Materials of Hebei Province, Research Center for Computational Physics, College of Physics Science and Technology, Hebei University, Baoding 071002, People's Republic of China

*E-mails: jlwang@hbu.edu.cn, shixq20hbu@hbu.edu.cn

**Abstract:** Momentum-matched type II van der Waals heterostructures (vdWHs) have been designed by assembling layered two-dimensional semiconductors (2DSs) with special band-structure combinations – that is, the valence band edge at the Γ point (the Brillouin-zone center) for one 2DS and the conduction band edge at the Γ point for the other [Ubrig *et al.*, Nat. Mater. 19, 299 (2020)]. However, the band offset sizes, band-alignment types, and whether momentum matched or not, all are affected by the interfacial effects between the component 2DSs, such as the quasichemical-bonding (QB) interaction between layers and the electrical dipole moment formed around the vdW interface. Here, based on density-functional theory calculations, first we probe the interfacial effects (including different QBs for valence and conduction bands, interface dipole, and, the synergistic effects of these two aspects) on band-edge evolution in energy and valley (location in the Brillouin zone) and the resulting changes in band alignment and momentum matching for a typical vdWH of monolayer InSe and bilayer $WS_2$, in which the band edges of subsystems satisfy the special band-structure combination for a momentum-matched type II vdWH. Then, based on the conclusions of the studied interfacial effects, we propose a practical screening method for robust momentum-matched type II vdWHs. This practical screening method can also be applied to other band alignment types. Our current study opens a way for practical screening and designing of vdWHs with robust momentum-matching and band alignment type.

# I. INTRODUCTION

Van der Waals (vdW) integration of two-dimensional semiconductors (2DSs) into heterostructures has received great attention for the several advantages of vdW heterostructures (vdWHs): high quality atomic-sharp interface due to being lattice mismatching-immune [1,2], diversity in material selection [3], controllable layer-number [4], and wide potential applications in electronics and optoelectronics. Many 2DSs can be used as components of vdWHs, such as transition metal dichalcogenides (TMDs) $MX_2$ (M = Mo, W; X = S, Se, Te) [5-8], group III chalcogenides MX (M = Ga, In; X = S, Se, Te) [9-12], group IV dichalcogenides $SnX_2$ (X = S, Se) [13], the $MA_2Z_4$ family (M = Mo, W; A = Si, Ge; Z = N, P, As) [14,15], layered oxide semiconductor (such as $Bi_2O_2Se$) [16] and Janus 2DSs [17,18], etc. The Anderson rule (AR) [19,20] is frequently used to understand or predict band alignments of a vdW heterostructure, which neglects the interlayer interaction and the charge redistribution at the interface of the component 2DSs. The band alignment from AR is then simplified to align the energy levels (relative to the vacuum levels of each system) of the free-standing 2DS components. Modifications to AR are necessary due to the interface dipole and the interfacial quasichemical-bonding interactions existing at the heterostructure interface, which not only change the energy of band edges of the component 2DSs, but also their valley [location in the Brillouin zone (BZ)] [19,21-23].

The band alignments of vdWHs can be classified into three main types: type I (symmetric), type II (staggered), and type III (broken). Each type has wide applications as devices [24-26]. Especially, the type II vdWHs enable efficient charge separation in different 2DS layers [27,28] and have been used as unipolar electronic devices, high electron mobility transistors, heterostructure solar cells, and field-effect phototransistors [29,30]. Momentum matching is important for high-efficiency photocarrier generation and transition [31]. Direct band gap 2DSs, if the band edges (including both valence and conduction bands) are not located at the Γ point of the BZ, are unlikely to be used for constructing robust momentum-matched heterostructures: as sketched in Fig. 1a taking the hexagonal BZ as an example. If the two BZs of two 2DSs are momentum-matched for a general *k*-point other than the center of BZ (Γ point), a random twist angle (θ) originated from the relative rotation between the two 2DS layers will then make momentum-mismatch [20,32-34]. In addition, the different lattice constants of two 2DSs (equivalent to rescaling the BZ of one 2DS relative to the other) may lead to momentum-mismatch if the vdWH is momentum-matched for a general *k*-point other than the Γ point. To solve this problem, in the year of 2020 Ubrig *et al.* proposed that, by designing heterostructure with valance band maximum (VBM) at the Γ point for one 2DS and the conduction band minimum (CBM) at the Γ point for the other, one can avoid any momentum-mismatch induced by rotational misalignment or lattice mismatch [32]. This method has the advantages of using indirect gap 2DSs as potential component-materials (Fig. 1b), which enables much more 2DSs as potential candidate materials. On the other hand, however, considering the effects at the vdW interface between 2DSs, such as the interface quasi-bonding interaction [22,35,36] and the interface dipole formed [21,37], both the *k*-space locations of band edges (Fig. 1b) and even the type of band alignment may change [38-40].

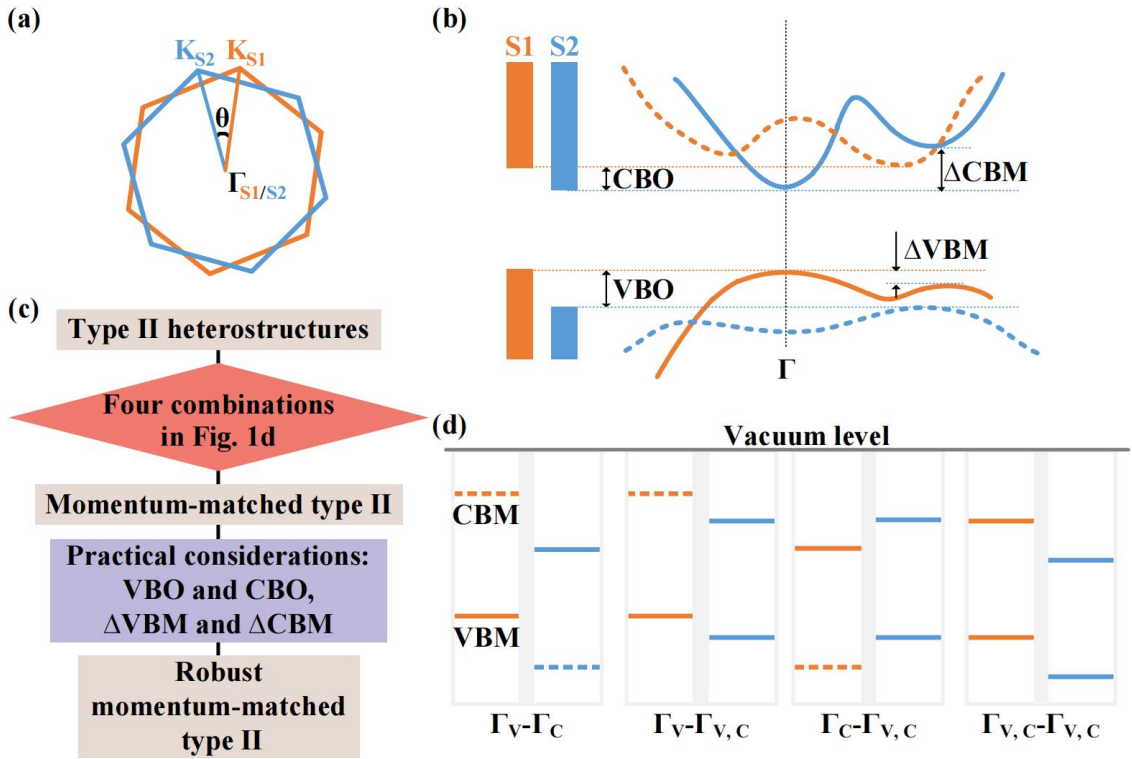

FIG. 1. Sketch maps for momentum-matching, band alignment, and the practical screening method for robust momentum-matched type II vdWHs from two 2DSs, S1 and S2. (a) Brillouin zones of S1 and S2 with a twist angle θ. (b) A typical example of the special band structure combination with VBM from S1 at the Γ point and CBM from S2 at the Γ point. VBO/CBO and ΔVBM/ΔCBM are correlated with the robustness of the band alignment type and momentum-matching, respectively. (c) The proposed practical screening method for robust momentum-matched type II vdWHs. (d) Four combinations of band edges for the screening. $\Gamma_V$, $\Gamma_C$, and $\Gamma_{V,C}$ denote the valence band edge, the conduction band edge, and both at the Γ point, respectively.

In the present work, firstly we systematically study the interfacial effects in a typical vdWH, including different quasi-bonding (QB) interactions in valence and conduction band edges, interface dipole, as well as the synergistic effects of these two factors that may affect the construction of a robust momentum-matched type II vdWH. In particular, the QB changes the valence and conduction band edges in a different manner, while the interface dipole shifts the band edges rigidly. In addition, QB has been proved to be related to orbital composition, symmetry and occupancy or not of the involved energy bands [23], and, the self-consistent potential and hence the interface dipole depends on the occupied QB states [41]. All these lead to various effects on the evolution of the band edge in vdWH relative to that of the subsystems [35,36,42]. Then, based on the insights from the study of a typical vdWH, we propose a practical screening method for robust type II heterostructures with robust momentum-matching (Fig. 1c). We find that, to keep robust band alignment type and moment-matching at the Γ point, two aspects in the band structure combinations need to be considered: 1) the magnitude of valence band offset (VBO) and conduction band offset (CBO), and 2) the minimum energy differences between valence and conduction band edges in different valleys (the ΔVBM and ΔCBM as labeled in Fig. 1b). In general, the band edges may locate at different valleys. Finally, the robustness of momentum-matched type II vdWHs can be obtained by the requirement that screening 2DSs with VBO and CBO to be larger than a certain value (~ 0.2 eV from our analysis), and, ΔVBM and ΔCBM to be larger than ~ 0.1 eV. This kind of screening methods, with constraints on the band structures of component subsystems (the band offset sizes and energy differences for band edges at different valleys), may thus also be applied to other types of band alignment, and, the rich database of 2DSs [26,43] provides great potential for screening momentum-matched vdWHs.

## II. CALCULATION METHODS

Density functional theory (DFT) [44] calculations were performed using the Vienna Ab initio Simulation Package (VASP) [45,46]. The projector augmented wave (PAW) potential [47,48] was adopted to describe the core electrons, and the valence electrons were described by plane wave basis with an energy cut-off of 500 eV. The exchange-correlation functional adopted the generalized gradient approximation (GGA) in the Perdew-Burke-Ernzerhof form (GGA-PBE) [49,50] for the analysis of interface effects. The vdW interaction in layered 2D materials was described by the DFT-

D3 method of Grimme *et al.* [51]. The InSe/bilayer WS$_2$ (BL-WS$_2$) heterostructure was built by stacking supercells of WS$_2$-($\sqrt{7} \times \sqrt{7}$)R19.1° and the InSe-(2×2). The corresponding Monkhorst-Pack *k*-point sampling was $5 \times 5 \times 1$ (or *k*-point density of $2\pi \times 0.03$ Å$^{-1}$). The lattice mismatch in the different vdWH models of the screened 26 robust momentum-matched type II vdWHs are calculated. Here, the space groups of the 2DS structure are given since the same chemical formula may have different structural phases, and the H-phase of BL-MX$_2$ ware used. The supercells used for the 26 vdWHs were summarized in Table SI in the Supplemental Material [52]. The vdWHs were modeled as slabs with a vacuum region of at least 15 Å to avoid the interaction between neighboring images. The convergence criteria were 10$^{-5}$ eV for the self-consistent energy calculation and 0.02 V/Å for the Hellmann-Feynman force in the geometric optimization. The VASPKIT program was used for post-processing of the electronic structure data [53]. The screening of vdWHs used the Computational 2D Materials Database (C2DB) [43,54]. To compare the relative strength of the interface interaction, we used the LOBSTER package [55] to obtain the *k*-resolved crystal orbital Hamilton population (COHP) [56].

## III. RESULTS AND DISCUSSION

To investigate the interfacial effects of a vdWH, 2D monolayer InSe and bilayer WS$_2$ (BL-WS$_2$) are adopted as a typical example since the band edges of the isolated monolayer InSe and BL-WS$_2$ meet the requirements of momentum-matching and type II band alignment [Figs. 2c(i) and 2c(iii)]. For InSe, the conduction band edge is located at Γ point, and, for BL-WS$_2$ the valence band edge is located at Γ point due to the interlayer QB interaction around Γ point [35]. The relaxed crystal structures of BL-WS$_2$, monolayer InSe, and InSe/BL-WS$_2$ vdWH are shown in Figs. 2a(i-iv). As a balance of computational cost and accuracy, supercells of InSe-(2 × 2) and BL-WS$_2$- ($\sqrt{7} \times \sqrt{7}$)R19.1° are used to construct the vdWH. To focus on the interfacial effects of QB and interface dipole, the same lattice constant is used in the calculations for subsystems with supercells [InSe-(2 × 2), BL-WS$_2$- ($\sqrt{7} \times \sqrt{7}$ )R19.1 ° ] and vdWH. Specifically, the averaged lattice-constant of subsystem supercells is adopted. To determine the interlayer separations of BL-WS$_2$ and InSe/BL-WS$_2$ vdWH, the binding energy curves are calculated as a function of interlayer separation (Fig. S1

[52]). The equilibrium distances of interlayer vertical-separation between WS$_2$ layers ($d_{S-S}$) and between InSe and WS$_2$ ($d_{S-Se}$) are 3.10 Å and 3.31 Å, respectively.

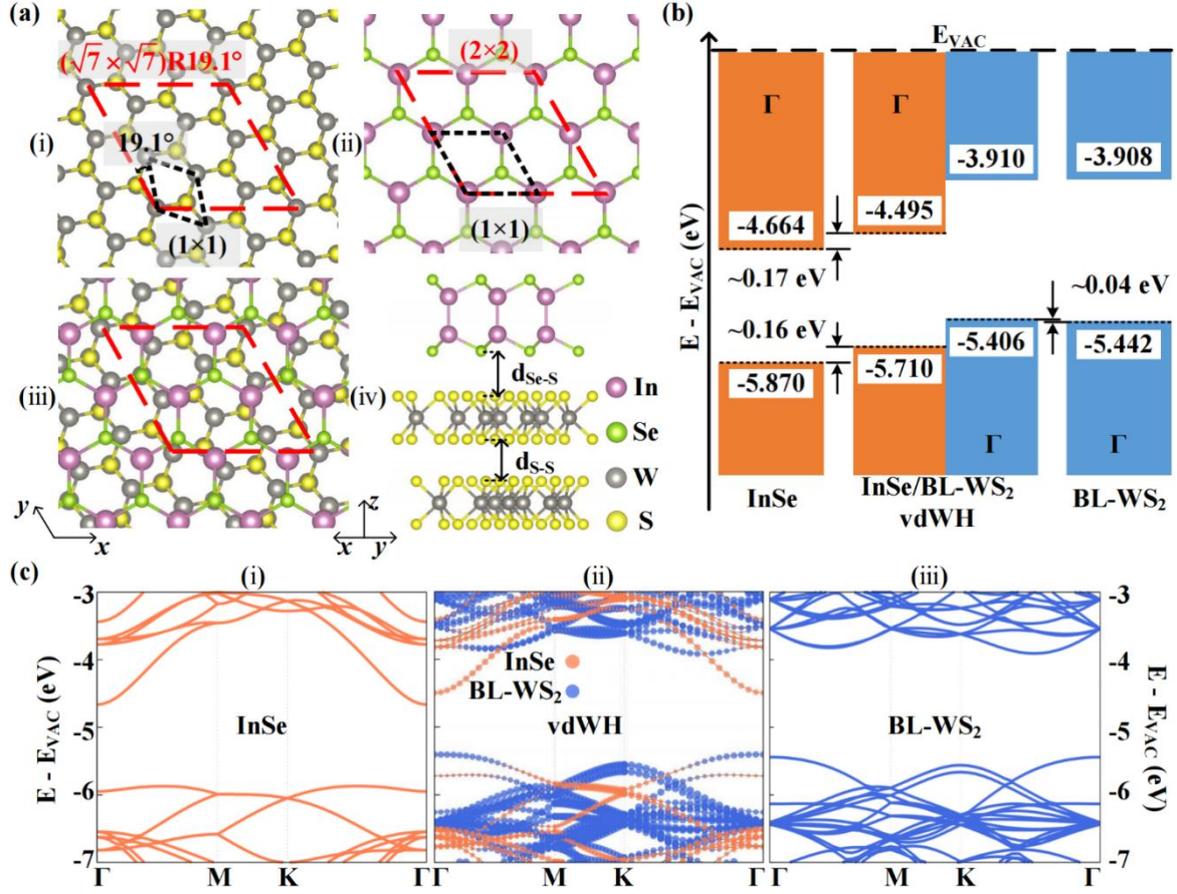

FIG. 2. (a) Geometric structure of InSe/BL-WS$_2$ vdWH: top-views of (i) BL-WS$_2$ with the (1×1) and ($\sqrt{7} \times \sqrt{7}$)R19.1° cells labeled out and (ii) InSe with (1×1) and (2×2) cells; (iii-iv) the top- and side-views of the InSe-(2×2)/BL-WS$_2$-($\sqrt{7} \times \sqrt{7}$)R19.1° vdWH. (b and c) The DFT calculated band alignment as shown with bars in (b) and with (projected) band structures in (c). The energy reference is the vacuum levels ($E_{VAC}$) of the isolated InSe, the InSe/BL-WS$_2$ vdWH, and the isolated BL-WS$_2$, respectively. In vdWH, there are two vacuum levels on the InSe and BL-WS$_2$ sides and the vacuum level on the BL-WS$_2$ side is set to zero.

The DFT calculated band alignment of the InSe/BL-WS$_2$ vdWH and that of their constituent 2DSs are shown in Fig. 2b. The labeled numbers in Fig. 2b are from the calculated band structures in Fig. 2c. Both InSe and BL-WS$_2$ are indirect bandgap semiconductors with energy gaps ($E_g$) of 1.21 eV and 1.53 eV, respectively. However, as shown in Figs. 2c(i) and 2c(iii), the conduction band edge of isolated InSe and the valence band edge of isolated BL-WS$_2$ are located at the Γ point. So, they are potential candidates for momentum-matched vdWH. Our heterostructure calculation proves that the InSe/BL-WS$_2$ vdWH has a direct band gap of 0.91 eV. The energy changes in VB (CB) from isolated subsystems to that in vdWH is 0.04 eV for BL-WS$_2$ (0.17 eV for InSe). The difference in energy change for VB and CB can be attributed to the different interface QB effects for BL-WS$_2$ and InSe (as will be explained in detail later). Fig. 2c(ii) shows the band structure of the InSe/BL-WS$_2$ vdWH projected to the subsystems. The valence band edge of the vdWH is mainly contributed by BL-WS$_2$ and the conduction band edge is mainly contributed by the InSe layer, indicating that electrons and holes can be separated with the formation of vdWH and the InSe/BL-WS$_2$ vdWH is a Γ-point momentum-matched type II heterostructure.

### A. Quasi-bonding interaction

Fig. 2c(ii) shows that the energy bands of monolayer InSe and BL-WS$_2$ are hybridized together for some valence bands, which is due to the orbital-hybridization at the interface and named as the covalent-like quasi-bonding (QB) interaction mainly from the out-of-plane orbitals of the surface atoms of vdW layers [22]. To gain deeper insights into the interface QB interaction between InSe and BL-WS$_2$, Fig. 3a shows the band structure projected to the out-of-plane $p_z$ orbitals of the Se and S atoms of the InSe/BL-WS$_2$ vdWH, and, the subsystem (InSe, monolayer and bilayer WS$_2$) band structures projected to $p_z$ and other orbitals as a reference can be find in Fig. S2 [52]. We focus on the valence bands in Fig. 3a since only the valence bands show significant interface QB interaction in Fig. 2c(ii). The necessary conditions for significant QB interaction between energy bands in adjacent vdW layers are: 1) the energy levels are close in energy and 2) the corresponding orbitals can overlap effectively, which require that they are close in real-space and both have out-of-plane orbitals [22,57]. From Figs. 2c(i) and 2c(iii), the valence bands at and around Γ point of the isolated 2DS components are close in energy (i.e., the valence band of InSe is between the two valence bands of BL-WS$_2$), which meets the first condition of QB interaction. Meanwhile, the Se and S atoms at the vdWH interface are close in real-space than the other elements [Fig. 2a(iv)] and the valence bands around Γ point have out-of-plane $p_z$ orbital for both Se and S atoms (refer to Fig. S2 [52]), which meets the second condition of QB interaction. Fig. 3a shows the band structure projected to the S-$p_z$ and Se-$p_z$ orbitals of the three valence-bands (VB, VB-1, and VB-2) at the Γ point for the InSe/BL-WS$_2$ vdWH. It is obvious that these three bands are coupled at and around the Γ point --- VB and VB-2 are mainly derived from the S-$p_z$ orbital of BL-WS$_2$, and VB-1 mainly from the Se-$p_z$ orbital of monolayer InSe.

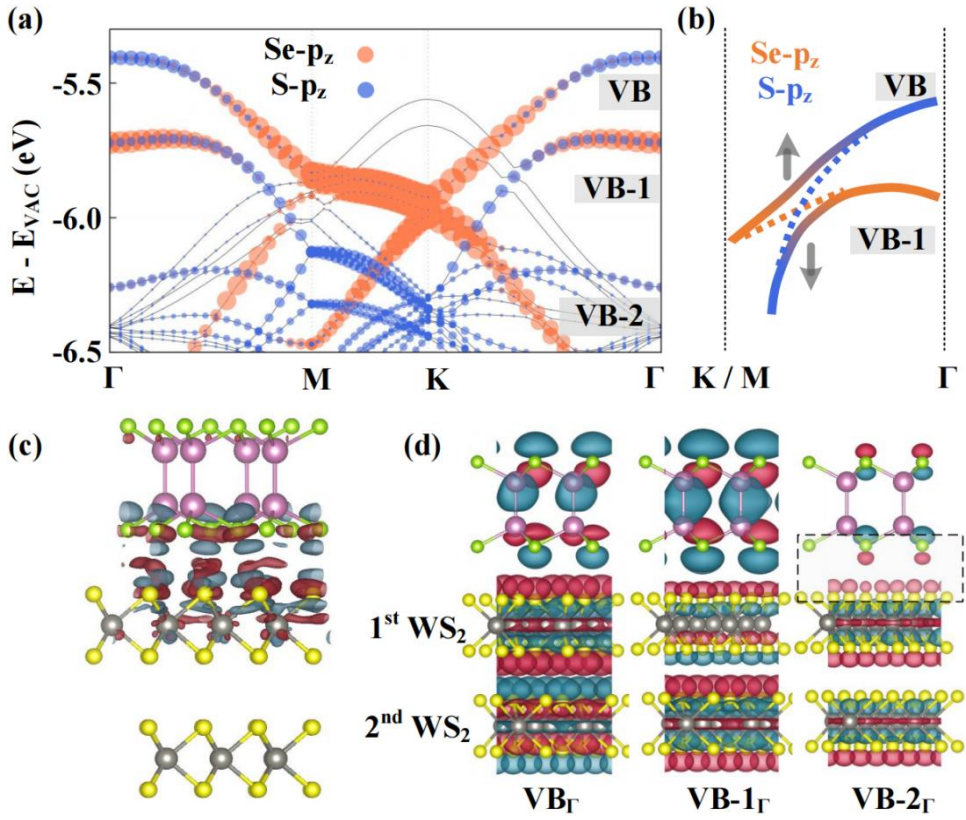

FIG. 3. (a) Band structure of the InSe/BL-WS$_2$ vdWH projected to the out-of-plane $p_z$ orbitals of the Se and S atoms. The dot sizes are proportional to the relative weight of contribution, and the grey lines indicate the full band structure of vdWH. (b) Sketch map of energy level splitting at the K-Γ (and M-Γ) path for the VB and VB-1 bands in (a). (c) Interface-QB-induced differential charge density of the InSe/BL-WS$_2$ (S---Se) interface. Blue and red denote electron depletion and accumulation, respectively. The isosurface value is $1.4\times10^{-4}$ e$^-$/a$_0^3$, where a$_0$ is the Bohr radius. (d) Wave-function plots at the Γ point corresponding to the three energy levels labeled in (a) with an isosurface of $6\times10^{-9}$ e$^-$/a$_0^3$. For VB-2$_\Gamma$ on the right, the wave-functions within the dashed rectangle indicate the interface bonding interaction; while the interface interaction is antibonding for VB$_\Gamma$.

For the WS$_2$ valence band at Γ, as shown in Fig. S3 [52], from monolayer to bilayer (BL), the energy increase of the antibonding state (of 0.50 eV) is greater than the energy decrease of bonding state (of 0.19 eV), and the energy splitting is 0.69 eV, similar to that in literature for MoS$_2$ [35]. When BL-WS$_2$ and InSe form the vdWH, due to the additional S---Se orbital hybridization at the interface of vdWH, the energy splitting between VB and VB-2 is increased to 0.85 eV (Fig. 3a) and the antibonding state (VB) shift up further by ~0.04 eV than that in BL-WS$_2$. This is one of the reasons for the band gap reduction in vdWH compared to that from the Anderson's rule. Fig. 3b sketches the ideal energy level splitting at the K-Γ (and M-Γ) path for the VB and VB-1 bands in Fig. 3a. Moreover, in Fig. 3c it can be seen from the differential charge density at the InSe/BL-WS$_2$ interface that electron accumulation occurs just in the middle of the vdW interface (also refer to Fig. 4a). The above interface interaction caused energy band further splitting and charge accumulation at the vdWH interface, both are the evidences for the interface QB interaction.

The wave-function plots at the Γ point for VB, VB-1, and VB-2 in Fig. 3d indicates that the orbital hybridization at the vdW interface is mainly contributed by the $p_z$ orbitals of the interface S and Se atoms, which is consistent with the projected band structure in Fig. 3a. It can be found that, at the vdWH interface, VB has antibonding character and VB-2 has bonding character. The VB-1 also has antibonding character but weaker than that in VB. To compare the relative strength of the interface QB interaction of the two antibonding states, the *k*-resolved crystal orbital Hamilton population (COHP) of the interface S-$p_z$---Se-$p_z$ pairs for VB at the Γ point, as listed in Table SII [52]. The orbital-projected COHP (*p*COHP) partitions the band-structure energy into the contribution of orbital-pair interactions, and the sum of *p*COHP can indicate the relative strength of interlayer quasi-bonding for our system. By summing the *p*COHP of the S---Se neighbors that the distance smaller than 4 Å at the interface in a supercell, we find that the sum of *p*COHP of VB at Γ point (VB@Γ) is 0.043, and that of VB-1 at Γ point (VB-1@Γ) is 0.028. This proved that the VB-1 also has antibonding character but weaker than that in VB. In contrast to the Γ point, at the K point in Fig. 3a, the first two valence bands (named as VB$_K$ and VB-1$_K$ from now on) are mainly contributed by the in-plane $p_x$ and $p_y$ orbitals of the S atoms (Fig. S2c [52]), which results in negligible interface QB with InSe and hence the energy of VB$_K$ and VB-1$_K$ almost does not change from free-standing BL-WS$_2$ to BL-WS$_2$ in vdWH in Figs. 2c(iii) and 2c(ii). In the analysis of the BL-WS$_2$ side of the vdWH, the rigid energy-shift from interface dipole can be neglected since for the InSe/BL-WS$_2$ vdWH the vacuum level at the WS$_2$ side was set to zero. Note that we only focus on the *p*-orbitals of S and Se atoms at the interface since these atoms are close to the vdW interface and hence dominate the QB interaction [22,57], while the orbitals of inner atoms such as the W-$d_{z^2}$ orbital are neglected.

The magnitude of energy level shift in valence band (VB) edge at Γ point due to interface QB, $\Delta_{QB}$, can be quantified as

$$\Delta_{QB} = VB_{\Gamma,\ vdWH} - VB_{\Gamma,\ BL-WS_2}, \#(1)$$

where vdWH means the InSe/BL-WS₂ heterostructure, and BL-WS₂ means free-standing BL-WS₂. We obtain $\Delta_{QB} \approx 0.04$ eV (as labeled in Fig. 2b) for QB interaction of the S---Se interface in vdWH. The magnitude of $\Delta_{QB}$ from isolated BL-WS₂ to BL-WS₂ in vdWH, ~0.04 eV, is smaller than that of MoS₂ from bilayer to trilayer (about 0.15 eV) [35], due to 1) the energy mismatch between energy levels from S of WS₂ and Se of InSe, and 2) the real-space mismatch between interface S and Se atoms in supercell (compared to the real-space matched S atoms of few-layer MoS₂ in unit cell). Note that the interface QB induced energy shift on the BL-WS₂ side, $\Delta_{QB}$(BL-WS₂), can be different to that on the InSe side, $\Delta_{QB}$(InSe). The $\Delta_{QB}$(InSe) is about 0.10 eV (but not simply ~0.16 eV as labeled in Fig. 2b), since the effect of the interface dipole should also be considered for the InSe side of vdWH. To analyze the band edge change on the InSe side, both the QB and the interface dipole should be considered.

## B. Interface dipole

The charge redistribution at the InSe and BL-WS₂ interface (Figs. 3c and 4a) results in the formation of an interface electric dipole and an interface potential step, $\Delta V$, across the vdWH interface. One way of visualizing the charge redistribution upon the formation of the InSe/BL-WS₂ (S---Se) interface is by the electron density difference (Fig. 3c)

$$\Delta\rho(x,y,z) = \rho_{vdWH} - \rho_{BL-WS_2} - \rho_{InSe}, \#(2)$$

where $\rho_{vdWH}$, $\rho_{BL-WS_2}$, and $\rho_{InSe}$ are the electron densities of the InSe/BL-WS₂ vdWH, isolated BL-WS₂ and InSe, respectively. By analyzing the (x, y) plane-averaged charge density, we can obtain the charge density difference along the direction normal to the interface, namely $\rho(z)$:

$$\rho(z) = \iint_A \rho(x,y,z)dxdy. \#(3)$$

Where $A$ is the interface area of the supercell used in the calculation model. And,

$$\Delta\rho(z) = \rho(z)_{vdWH} - \rho(z)_{BL-WS_2} - \rho(z)_{InSe}. \#(4)$$

The $\Delta\rho(z)$ is mainly localized around the InSe/BL-WS₂ (S---Se) interface. Fig. 4a shows the plane-averaged charge density, $\Delta\rho(z)$, along the z-direction.

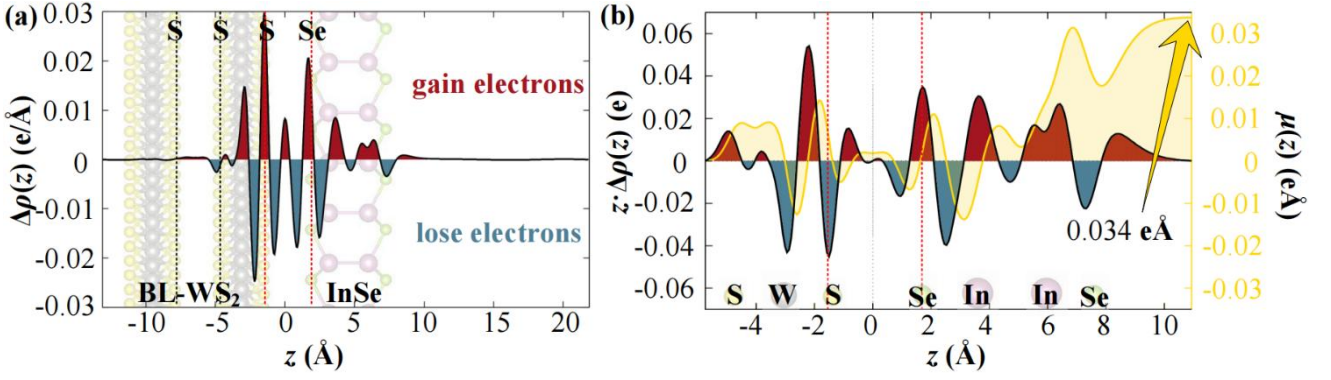

FIG. 4. (a) The electron density difference along the $z$-direction normal to the interface of the InSe/BL-WS$_2$ vdWH. The WS$_2$/InSe (S---Se) interface region is indicated with two red dotted lines. The WS$_2$-WS$_2$ (S---S) interlayer region is shown with black dotted lines. (b) The $z \times \Delta\rho(z)$ and the dipole moment, $\mu(z)$, around the S---Se interface.

The interfacial dipole moment, $\mu(z)$, can be calculated from $\Delta\rho(z)$ using the following formula:

$$\mu(z) = \int_{-5.74}^{10.94} z\Delta\rho(z)dz. \#(5)$$

Here, the upper and lower limits of the integration is determined by the decay of the differential charge density, $\Delta\rho(z)$ in Fig. 4, as it goes away from the vdW interface. At the InSe side, it decays to zero in the vacuum, which gives the upper limit of 10.94 Å. For the WS$_2$ layer near the interface, $\Delta\rho(z)$ decays to zero in the vdW gap of bilayer WS$_2$, which gives the lower limit of -5.74 Å. And, the origin of the vdW interface ($z = 0$ in Fig. 4a) is set to a point where the charge accumulation reaches maximum due to the covalent-like quasi-bonding character of the interface. In literature [58,59], the origin ($z = 0$) was set to a point where the charge accumulation switches to charge depletion. We find that our definition from quasi-bonding also satisfies 1) the charge accumulation switches to charge depletion, namely, the summed charge (or integration of $\Delta\rho(z)$) is zero at the defined origin ($z = 0$), as shown in Fig. S4 [52]; and, 2) close to the medium of vdW gap and a little closer to sulfur side, which is consistent with the fact that the radius of sulfur atom is smaller than selenium atom. For the integration of $\Delta\rho(z)$, the lower limit of integration is given in Eq. 5. A dipole moment $\mu(z)$ of +0.163 Debye (or 0.034 eÅ) is found, thus confirming the formation of a vdW interface-induced electric dipole moment (By Eq. 7, the potential step ($\Delta V$) can be obtained from $\mu(z)$).

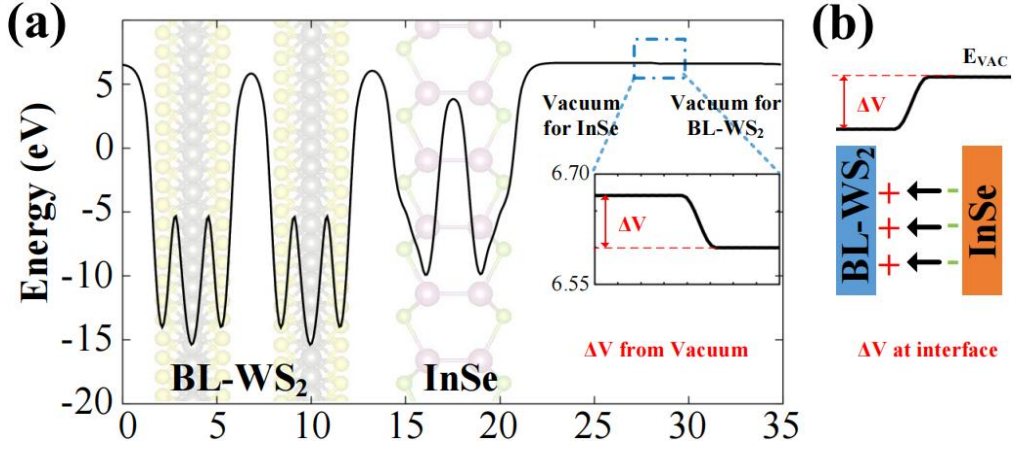

FIG. 5. (a) Plane-averaged electrostatic potential and the interface potential step, ΔV, for the InSe/BL-WS$_2$ vdWH. The interface potential step ΔV, which can be obtained from the difference between the vacuum levels of InSe side and BL-WS$_2$ side, is shown in the inset. (b) The schematic illustration of the ΔV and the interface dipole at the interface.

The magnitude of the electrostatic potential step, $\Delta V$, at the interface can be used to quantify the effect of interface dipole, which can also be obtained from plane-averaged electrostatic potential from DFT calculations (Fig. 5) with a dipole correction [60]. Here, $\Delta V$ is shown in the inset of Fig. 5a in the vacuum and in Fig. 5b for the BL-WS$_2$/InSe interface. Note the sign reversing of $\Delta V$ from vacuum to that at the interface, since $\Delta V = E_{vac}(\text{InSe}) - E_{vac}(\text{BL-WS}_2)$ [60]. The magnitude of $\Delta V$ is 0.071 eV for the InSe/BL-WS$_2$ interface. Both the positive value of $\mu(z)$ (Fig. 4b) and $\Delta V$ (Fig. 5) suggest an interface dipole points from InSe layer (with electrons) to the BL-WS$_2$ (with holes), as shown in the bottom panel in Fig. 5b. Such a dipole rigidly shifts the band structure (both VB and CB) of InSe in the InSe/BL-WS$_2$ heterostructure up relative to that of BL-WS$_2$. The relative movement of CB and VB reduces the VBO and CBO by ~0.07 eV due to interface dipole. It is reported that the interface dipole is largely depends on VBO: when VBO is small, the interface dipole is small [21].

One can also obtain $\Delta V$ by solving the one-dimensional Poisson equation [61]:

$$\nabla^2 V(z) = -\frac{\Delta \rho(z)}{\varepsilon_0}, \quad \#(6)$$

where $V(z)$ is the electrostatic potential, $\varepsilon_0$ is the vacuum permittivity. Solving the Poisson equation with $\Delta \rho(z)$ as source then gives a potential step across the interface [60]:

$$\Delta V = \frac{e^2}{\varepsilon_0 A} \int z \Delta \rho(z)\, dz \approx \frac{e^2}{\varepsilon_0 A} \mu(z), \#(7)$$

where $e$ is the (positive) elementary charge. From the above equation, one can see that $\Delta V$ is positively correlated with $\mu(z)$. The calculated $\Delta V$ from Eq. 7 has a similar value as the above value obtained from vacuum levels.

The other factor, QB, on band alignment could be more complicated as detailed below.

### C. Quantifying the $\Delta_{QB}$s for VB and CB

Due to the interface QB interaction, the VB of InSe and the two VBs of BL-WS$_2$ are coupled at the $\Gamma$ point which generates three VBs in the InSe/BL-WS$_2$ vdWH (Fig. 6a, also refer to Figs. 2c(ii) and 3a). Fig. S2a [52] shows that the VB and CB at the $\Gamma$ point (VB$_\Gamma$ and CB$_\Gamma$) of InSe have similar orbital components (mainly the In-$s$ and Se-$p_z$ orbitals). Also, as shown in Fig. 6b, the VB$_\Gamma$ and CB$_\Gamma$ are bonding and antibonding states of the intralayer In-$s$ orbitals. Through the mediation of the intralayer In-$s$ and Se-$p_z$ orbitals that both VB and CB have, the up-shift of CB$_\Gamma$ occurs following the up-shift of VB-1$_\Gamma$ (see the grey arrow in Fig. 6a), and the up-shift of VB-1$_\Gamma$ in vdWH (relative to VB$_\Gamma$ of isolated InSe) are from the interface QB in VBs as discussed above. So, in Fig. 2b the energy up-shift of the InSe CB edge (~0.17 eV) is very close to that of the InSe VB edge (~0.16 eV), although there is no apparent QB for the CB edge. Recall that the energy shift of band edges in vdWH mainly comes from two factors: interface QB and interface dipole. Recall that the interface dipole between InSe and BL-WS$_2$ is ~0.07 eV. Therefore, the increase of InSe VB (and CB) due to QB is ~0.1 eV [0.16 eV for VB (or 0.17 eV for CB) in total minus 0.07 eV from interface dipole].

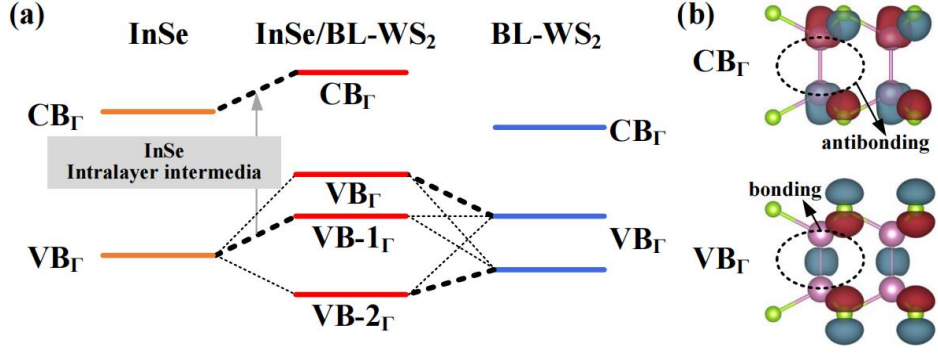

FIG. 6. Origin of the InSe CB up-shift from InSe subsystem to that in vdWH (refer to Figs. 2b and 2c). (a) The dashed lines represent the QB interaction between energy levels of subsystems, and the bold dashed lines indicate the levels with a major contribution from subsystem to vdWH. The vertical grey-arrow indicates that the up-shift of CB is intermediated by the InSe intralayer orbitals. (b) The wave functions of monolayer InSe for VB and CB at the $\Gamma$ point, which are bonding and antibonding states of the intralayer In-$s$ orbitals.

Both interface QB and interface dipole affects the band offsets, VBO and CBO, which are given by:

$$\text{VBO} = \text{VBO}_{AR} + \Delta V + \Delta_{QB}(\text{VB}), \quad (8)$$

$$\text{CBO} = \text{CBO}_{AR} + \Delta V + \Delta_{QB}(\text{CB}) \quad (9)$$

where the subscript AR means the band offset of isolated subsystems given by the Anderson's rule. $\Delta_{QB}(\text{VB})$ and $\Delta_{QB}(\text{CB})$ are the modification to band offsets by interface QB interaction at VB and CB, respectively. For the InSe/BL-WS$_2$ vdWH:

$$\Delta_{QB}(\text{VB}) = \Delta_{QB}(\text{InSe}) - \Delta_{QB}(\text{BL-WS}_2) \approx 0.10 \text{ eV} - 0.04 \text{ eV} = 0.06 \text{ eV}, \text{ and}$$

$$\Delta_{QB}(\text{CB}) = \Delta_{QB}(\text{InSe}) \approx 0.10 \text{ eV}.$$

As mentioned above, $\Delta V$ is ~0.07 eV. Then, the DFT calculated VBO and CBO in Fig. 2b are understood well from Eqs. (8-9), which quantify the partial contributions of interface dipole and interface QB.

## D. Practical screening method

From subsystems to vdWH for the above typical example, the interlayer QB interaction in our system gives a correction in band offsets (VBO and CBO) range from 0.06 eV [for $\Delta_{QB}(VB)$ discussed above] to 0.1 eV [for $\Delta_{QB}(CB)$], which is expected to be similar for the correction to $\Delta CBM$ and $\Delta VBM$ in Fig. 1b. In addition, the energy shifts due to $\Delta V$ is ~0.07 eV, which only affect the band offsets. Therefore, both QB and interface dipole affect VBO and CBO (the maximum effect is the addition of the maximum absolutions values of both factors, i.e., 0.1 eV + 0.07 eV ≈ 0.2 eV), while only QB affects $\Delta CBM$ and $\Delta VBM$ (~0.1 eV). Thus, for the screening of a robust momentum-matched type II vdWH, a simple and reasonable deduction is that the magnitude of VBO and CBO from free-standing subsystems to be both larger than ~0.2 eV, and, $\Delta VBM$ and $\Delta CBM$ both larger than ~0.1 eV. For the effect of QB, this should be enough for thicker 2DSs more than monolayer since the additional QB effects (band edge up-shift) from the vdWH interface become smaller as the 2DS becomes thicker [35]. For the effects of the interface dipole in type II vdWH, this should also be sufficient, since the $\Delta V$ in vdWH has been demonstrated to be smaller than 0.1 eV for VBO smaller than ~0.7 eV [21].

Actually, for vdWHs composed of two monolayer 2DSs, for example, for the HfSe$_2$ (P-4m2)/InS (P-6m2) vdWH, our calculation shows that the above criterion is also satisfied. This is due to, in general, in vdWH there are 1) energy level mismatch in the two 2DSs, 2) valley mismatch (different *k*-point in BZ) between band edges of the two 2DSs, and 3) real-space mismatch for interface atoms in a supercell (lattice constant mismatch between 2DSs induces real-space local mismatch) which may increase the separation of the vdW gap at the vdWH interface. All these factors weaken the interface interaction.

Based on the above discussions, we propose a practical screening method for robust type II vdWH with robust momentum-matching. Some monolayer 2DSs from the Computational 2D Materials Database (C2DB) [43,54] and some our calculated bilayer-2DSs were used for the screening. Figs. 7 and 8 show the screening method for robust momentum-matched vdWH.

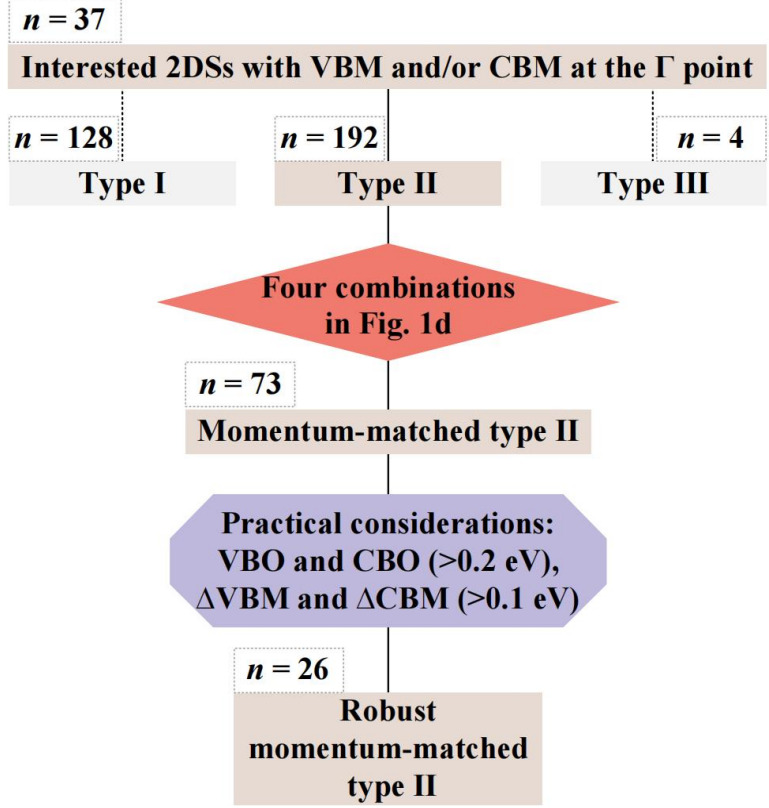

FIG. 7. Screening of robust momentum-matched vdWH with robust type II band alignment from interested 2DSs with VBM and/or CBM at the Γ point. $n$ is the number of structures considered or screened out.

Fig. 7 shows that, out of 37 2DSs with VBM and/or CBM at the Γ point, there are 128 type I vdWHs, 192 type II vdWHs, and 4 type III vdWHs. For the convenience of screening type II vdWHs, 2DSs with VBM and/or CBM at the Γ point are classified into three categories: only VBM at the Γ point ($Γ_V$), only CBM at the Γ point ($Γ_C$), and both at the Γ point ($Γ_{V,C}$). From these three categories, four combinations of 2DS band edges for the construction of momentum-matched type II vdWHs are summarized in Fig. 1d and Fig. 8: namely, $Γ_V$ - $Γ_C$, $Γ_V$ - $Γ_{V,C}$, $Γ_C$ - $Γ_{V,C}$, and $Γ_{V,C}$ - $Γ_{V,C}$. Based on the Anderson's rule, there are 73 momentum-matched type II vdWHs. Then, for the robustness of band alignment type and moment-matching, a constraining of VBO and CBO of free-standing subsystems to be both larger than 0.2 eV, and, ΔCBM and ΔVBM larger than 0.1 eV are considered. Based on the above considerations, there are 26 robust momentum-matched type II vdWHs.

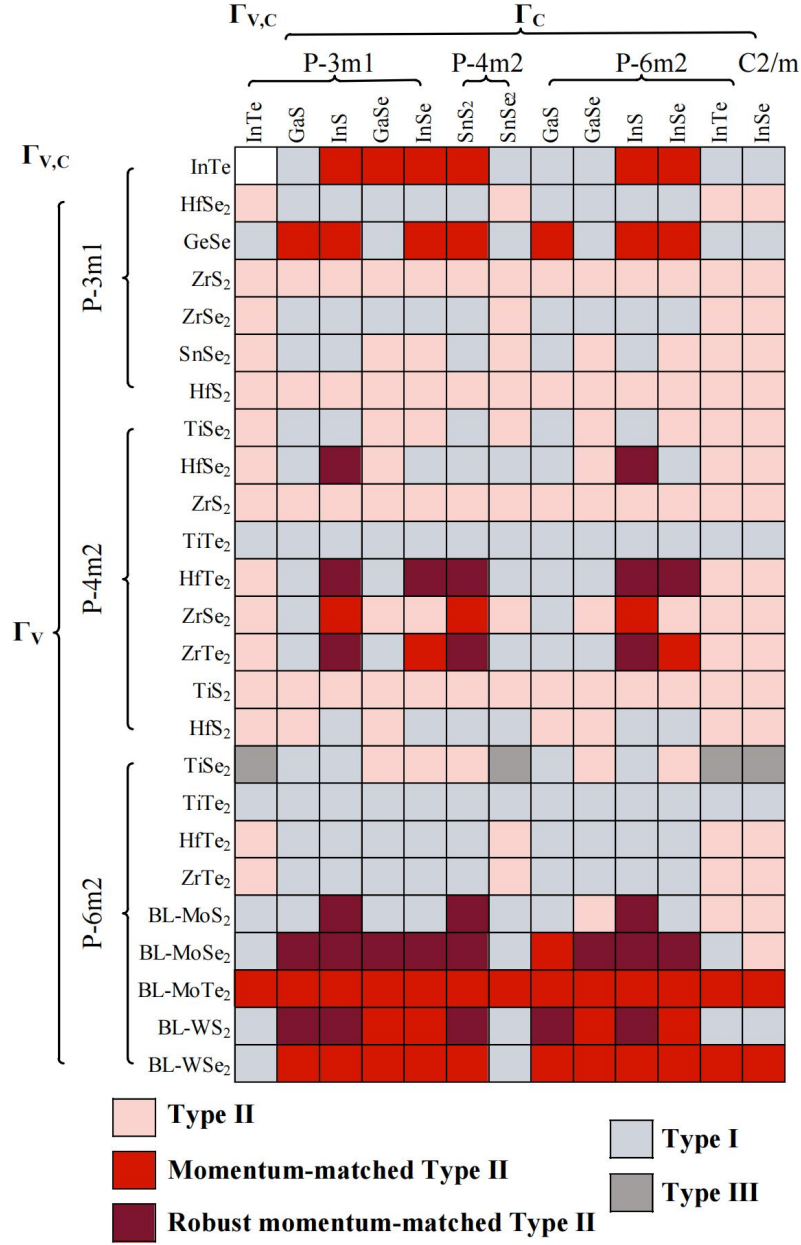

FIG. 8. Screening of momentum-matched type II vdWHs from some typical 2DSs with CBM and/or VBM at the Γ point. $\Gamma_V$, $\Gamma_C$, and $\Gamma_{V,C}$ denote the valence band edge, the conduction band edge, and both at the Γ point, respectively. The space groups are given since the same chemical formula may have different structural phases. For bilayers (BL), the H-phase were used.

Fig. 8 shows the vdWHs from the 37 interested monolayer and bilayer 2DSs with VBM and/or CBM at the Γ point, including (mainly) transition metal dichalcogenides (TMDs, or $TMX_2$), group III chalcogenides (MX, actually $M_2X_2$), and group IV dichalcogenides ($MX_2$), etc. The space groups are also given since the same chemical formula may have different structural phases. Some bilayer TMDs are included since they have valence band edge at the Γ point.

Fig. 8 shows that, among the 37 2DSs with VBM and/or CBM at the Γ point, 24 have VBM at the Γ point ($\Gamma_V$ in Fig. 8), 12 have CBM at the Γ point ($\Gamma_C$), and one (InTe) has both VBM and CBM at the Γ point ($\Gamma_{V, C}$). The band edges relative to the vacuum level of isolated subsystem of monolayer 2DSs used the results from the C2DB [43,54] and that of the bilayers were from our calculations, as summarized in Table SIII in the Supplemental Material [52]. Fig. 8 shows the constructed vdWHs including type I, type II, and type III heterostructures according to the AR. The different types are denoted by different colors. Type II vdWHs are further classified into subtypes. Out of 192 type II vdWHs, there are 73 momentum-matched ones and 26 "robust" momentum-matched ones. To check the accuracy of our proposed practical screening method, the screened 26 robust momentum-matched type II vdWHs are calculated. As shown in Fig. S5 [52], all 26 heterostructures are momentum-matched type II vdWHs. The supercells used for the simulated 26 robust momentum-matched type II vdWHs and the corresponding lattice mismatch are listed in the Table SI [52]. If higher quality input data are available (such as from calculations/databases of more advanced density-functional or even from the experimental data), they can be used as the input data for the screening method. In addition, the different vdW correction methods have influences on the alignment of the band edges of layered 2D semiconductors, since the vdW dispersion interactions change the interlayer separation [62-64] and hence affect the interface QB. Since the DFT+D3 method may underestimate the distance between layers, more advanced many-body dispersion energy method (MBD) corrections [65] increase the interlayer separation [62]. We calculated the interlayer distance and the corresponding band structure with the PBE+MBD method. As shown in Figs. S6 and S7, the MBD corrections increase the interlayer separation by 0.08 Å than the D3 corrections, and the band alignment type does not change. Compared to PBE+MBD, since our adopted PBE+D3 method may underestimates the interlayer separation and hence overestimates the interface QB interaction. This means that if the screened robust momentum-matched type II vdWHs passed the check of the PBE+D3 calculation, they could also pass the PBE+MBD calculation.

## IV. CONCLUSION REMARKS

In conclusion, a practical screening method is proposed for the constructing of robust momentum-matched type II heterostructures, considering the energy band changes induced by interfacial effects, including interface QB interaction and interface dipole. The influence of these two interactions on the energy band alignment and momentum-matching is studied using a typical vdWH composed of monolayer InSe and BL-WS$_2$. The interface dipole moves CB and VB rigidly, while the effect of interface QB is energy and valley dependent. We find that the effect of both factors, the additional QB interaction at the vdWH interface and the interface dipole, on the band offset and the ΔVBM/ΔCBM is small in the vdWH. The practical screening method could be extended to other band alignment types, and useful for identifying the critical features in a machine learning model.

In the above, the screening method is demonstrated with the data (band edges) from the PBE functional as input for the screening, the screening method itself does not depend on the quality of the input data, although the final screened vdWHs may change. If higher quality input data are available (such as from calculations/databases of more advanced density-functional or even from the experimental data), they can be used as the input data for the screening method. This work mainly focuses on the interface QB and the interface dipole formed at the vdWH interface, and other factors, such as the interface exciton induced gap reduction [66], may be added by modifying the constraint(s) of VBO and CBO, and/or ΔCBM and ΔVBM. In addition, other constraints maybe added, for example, for the visible light spectrum of interlayer transition, one can choose the vdWHs with suitable band gaps (between 1.68 and 3.26 eV) [67,68].

## Acknowledgments

This work was supported by the Natural Science Foundation of China (Grants Nos. 12274111 and 12104124), the Natural Science Foundation of Hebei Province of China (Grants Nos. A2021201001 and A2021201008), the Postgraduate's Innovation Fund Project of Hebei University (Grant No. HBU2023BS006), the Advanced Talents Incubation Program of the Hebei University (Grants Nos. 521000981390, 521000981394, 521000981395, 521000981423, and 521100221055), the Scientific Research and Innovation Team of Hebei University (No. IT2023B03), and the high-performance computing center of Hebei University.

*Supplemental Materials for:*

# Momentum-matching and band alignment type in van der Waals heterostructures: Interfacial effects and materials screening

Yue-Jiao Zhang, Yin-Ti Ren, Xiao-Huan Lv, Xiao-Lin Zhao, Rui Yang, Nie-Wei Wang, Chen-Dong Jin, Hu Zhang, Ru-Qian Lian, Peng-Lai Gong, Rui-Ning Wang, Jiang-Long Wang*, and Xing-Qiang Shi*

TABLE SI. Supercells used for the simulated 26 robust momentum-matched type II vdWHs and the corresponding lattice mismatch along the two in-plane directions.

| vdWH | Supercell | Lattice mismatch |
|---|---|---|
| HfSe$_2$ (P-4m2)/InS (P-3m1) | $\begin{pmatrix}1 & 0\\ 6 & 5\end{pmatrix}/\begin{pmatrix}6 & -3\\ 0 & 1\end{pmatrix}$ | 0.01%, 0.31% |
| HfSe$_2$ (P-4m2)/InS (P-6m2) | $\begin{pmatrix}5 & 5\\ 0 & 1\end{pmatrix}/\begin{pmatrix}8 & 2\\ 1 & 1\end{pmatrix}$ | 0.08%, 0.9% |
| HfTe$_2$ (P-4m2)/InS (P-3m1) | $\begin{pmatrix}5 & 4\\ -1 & 3\end{pmatrix}/\begin{pmatrix}4 & 2\\ -3 & 5\end{pmatrix}$ | 0.46%, 0.04% |
| HfTe$_2$ (P-4m2)/InS (P-6m2) | $\begin{pmatrix}1 & 5\\ -3 & 1\end{pmatrix}/\begin{pmatrix}5 & 6\\ -2 & 2\end{pmatrix}$ | 0.31%, 0.25% |
| HfTe$_2$ (P-4m2)/InSe (P-3m1) | $\begin{pmatrix}2 & 2\\ 0 & -6\end{pmatrix}/\begin{pmatrix}3 & 0\\ -2 & 5\end{pmatrix}$ | 0.04%, 0.00% |
| HfTe2 (P-4m2)/InSe (P-6m2) | $\begin{pmatrix}6 & 0\\ 2 & -2\end{pmatrix}/\begin{pmatrix}5 & -2\\ 3 & 0\end{pmatrix}$ | 0.14%, 0.21% |
| HfTe$_2$ (P-4m2)/SnS$_2$ (P-4m2) | $\begin{pmatrix}2 & 0\\ 0 & 2\end{pmatrix}/\begin{pmatrix}2 & 1\\ -1 & 2\end{pmatrix}$ | 0.34%, 0.34% |
| ZrTe$_2$ (P-4m2)/InS (P-3m1) | $\begin{pmatrix}3 & 1\\ -1 & 5\end{pmatrix}/\begin{pmatrix}6 & 5\\ -2 & 2\end{pmatrix}$ | 0.06%, 0.12% |
| ZrTe$_2$ (P-4m2)/InS (P-6m2) | $\begin{pmatrix}1 & 5\\ -3 & 1\end{pmatrix}/\begin{pmatrix}5 & 6\\ -2 & 2\end{pmatrix}$ | 0.15%, 0.21% |
| ZrTe$_2$ (P-4m2)/SnS$_2$ (P-4m2) | $\begin{pmatrix}2 & 3\\ -3 & 2\end{pmatrix}/\begin{pmatrix}1 & 4\\ -4 & 1\end{pmatrix}$ | 0.32%, 0.32% |
| BL-MoS$_2$/InS (P-3m1) | $\begin{pmatrix}5 & 0\\ 0 & 5\end{pmatrix}/\begin{pmatrix}4 & 0\\ 0 & 4\end{pmatrix}$ | 0.71%, 0.71% |
| BL-MoS$_2$/InS (P-6m2) | $\begin{pmatrix}5 & 0\\ 0 & 5\end{pmatrix}/\begin{pmatrix}4 & 0\\ 0 & 4\end{pmatrix}$ | 1.20%, 1.20% |
| BL-MoS$_2$/SnS$_2$ (P-4m2) | $\begin{pmatrix}8 & 6\\ 1 & 3\end{pmatrix}/\begin{pmatrix}6 & 1\\ 1 & 2\end{pmatrix}$ | 0.01%, 0.41% |
| BL-MoSe$_2$/InS (P-3m1) | $\begin{pmatrix} 6 & 3 \\ -3 & 3 \end{pmatrix} / \begin{pmatrix} 5 & 2 \\ -2 & 3 \end{pmatrix}$ | 0.38%, 0.38% |
| BL-MoSe$_2$/InS (P-6m2) | $\begin{pmatrix} 6 & 3 \\ 3 & 6 \end{pmatrix} / \begin{pmatrix} 5 & 2 \\ 3 & 5 \end{pmatrix}$ | 0.47%, 0.47% |
| BL-MoSe$_2$/GaSe (P-3m1) | $\begin{pmatrix} 2 & 2 \\ 0 & 2 \end{pmatrix} / \begin{pmatrix} 2 & 1 \\ 1 & 2 \end{pmatrix}$ | 0.07%, 0.07% |
| BL-MoSe$_2$/GaSe (P-6m2) | $\begin{pmatrix} 2 & 2 \\ 0 & 2 \end{pmatrix} / \begin{pmatrix} 2 & 1 \\ 1 & 2 \end{pmatrix}$ | 0.12%, 0.12% |
| BL-MoSe$_2$/InSe (P-3m1) | $\begin{pmatrix} 5 & 0 \\ 0 & 5 \end{pmatrix} / \begin{pmatrix} 4 & 0 \\ 0 & 4 \end{pmatrix}$ | 0.72%, 0.72% |
| BL-MoSe$_2$/InSe (P-6m2) | $\begin{pmatrix} 6 & 2 \\ -2 & 4 \end{pmatrix} / \begin{pmatrix} 5 & 2 \\ -2 & 3 \end{pmatrix}$ | 0.64%, 0.64% |
| BL-MoSe$_2$/GaS (P-3m1) | $\begin{pmatrix} 4 & 4 \\ 0 & 4 \end{pmatrix} / \begin{pmatrix} 4 & 3 \\ 1 & 4 \end{pmatrix}$ | 0.50%, 0.50% |
| BL-MoSe2/SnS$_2$ (P-4m2) | $\begin{pmatrix} 4 & 1 \\ 1 & 4 \end{pmatrix} / \begin{pmatrix} 3 & 1 \\ -1 & 3 \end{pmatrix}$ | 0.16%, 0.09% |
| BL-WS$_2$/InS (P-3m1) | $\begin{pmatrix} 3 & 5 \\ -2 & 3 \end{pmatrix} / \begin{pmatrix} 1 & 4 \\ -3 & 1 \end{pmatrix}$ | 0.82%, 0.81% |
| BL-WS$_2$/InS (P-6m2) | $\begin{pmatrix} 5 & 0 \\ 0 & 5 \end{pmatrix} / \begin{pmatrix} 4 & 0 \\ 0 & 4 \end{pmatrix}$ | 0.95%, 0.95% |
| BL-WS$_2$/SnS$_2$ (P-4m2) | $\begin{pmatrix} 3 & 1 \\ -3 & 5 \end{pmatrix} / \begin{pmatrix} 2 & 1 \\ -5 & 3 \end{pmatrix}$ | 0.76%, 0.08% |
| BL-WS$_2$/GaS (P-3m1) | $\begin{pmatrix} 2 & 2 \\ 0 & 2 \end{pmatrix} / \begin{pmatrix} 2 & 1 \\ 1 & 2 \end{pmatrix}$ | 0.60%, 0.60% |
| BL-WS$_2$/GaS (P-6m2) | $\begin{pmatrix} 2 & 2 \\ 0 & 2 \end{pmatrix} / \begin{pmatrix} 2 & 1 \\ 1 & 2 \end{pmatrix}$ | 0.75%, 0.75% |

The primitive cell and supercell lattice vectors are defined as $\begin{pmatrix} \vec{a} \\ \vec{b} \end{pmatrix}$ and $\begin{pmatrix} \vec{A} \\ \vec{B} \end{pmatrix}$, respectively. The transformation matrix is $\begin{pmatrix} m_{11} & m_{12} \\ m_{21} & m_{22} \end{pmatrix}$, and the transformation formula between the primitive cell and supercell lattice vectors is given by the following matrix product form: $\begin{pmatrix} \vec{A} \\ \vec{B} \end{pmatrix} = \begin{pmatrix} m_{11} & m_{12} \\ m_{21} & m_{22} \end{pmatrix} \begin{pmatrix} \vec{a} \\ \vec{b} \end{pmatrix}$. For instance, the transformation matrix $\begin{pmatrix} 5 & 5 \\ 0 & 1 \end{pmatrix}$ and $\begin{pmatrix} 8 & 2 \\ 1 & 1 \end{pmatrix}$ corresponding to supercells HfSe$_2$ (P-4m2) - ($5\sqrt{2} \times 1$) and InS (P-6m2) - ($\sqrt{52} \times 1$), respectively. The lattice mismatch in HfSe$_2$ (P-4m2)/InS (P-6m2) heterostructure along $\vec{A}$ and $\vec{B}$ are 0.08% and 0.9%, respectively.

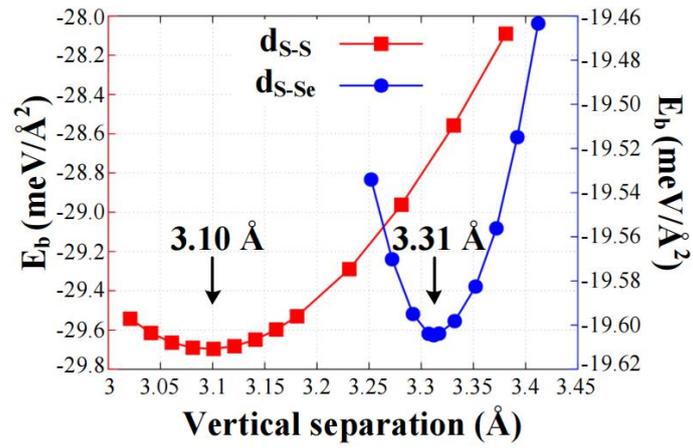

FIG. S1 Binding energy curves as a function of the interlayer distance for the BL-WS$_2$ ($d_{S-S}$, red curve) and for the InSe/BL-WS$_2$ heterostructure ($d_{S-Se}$, blue curve). See Fig. 2a(iv) in the main text for $d_{S-S}$ and $d_{S-Se}$ in the heterostructure.

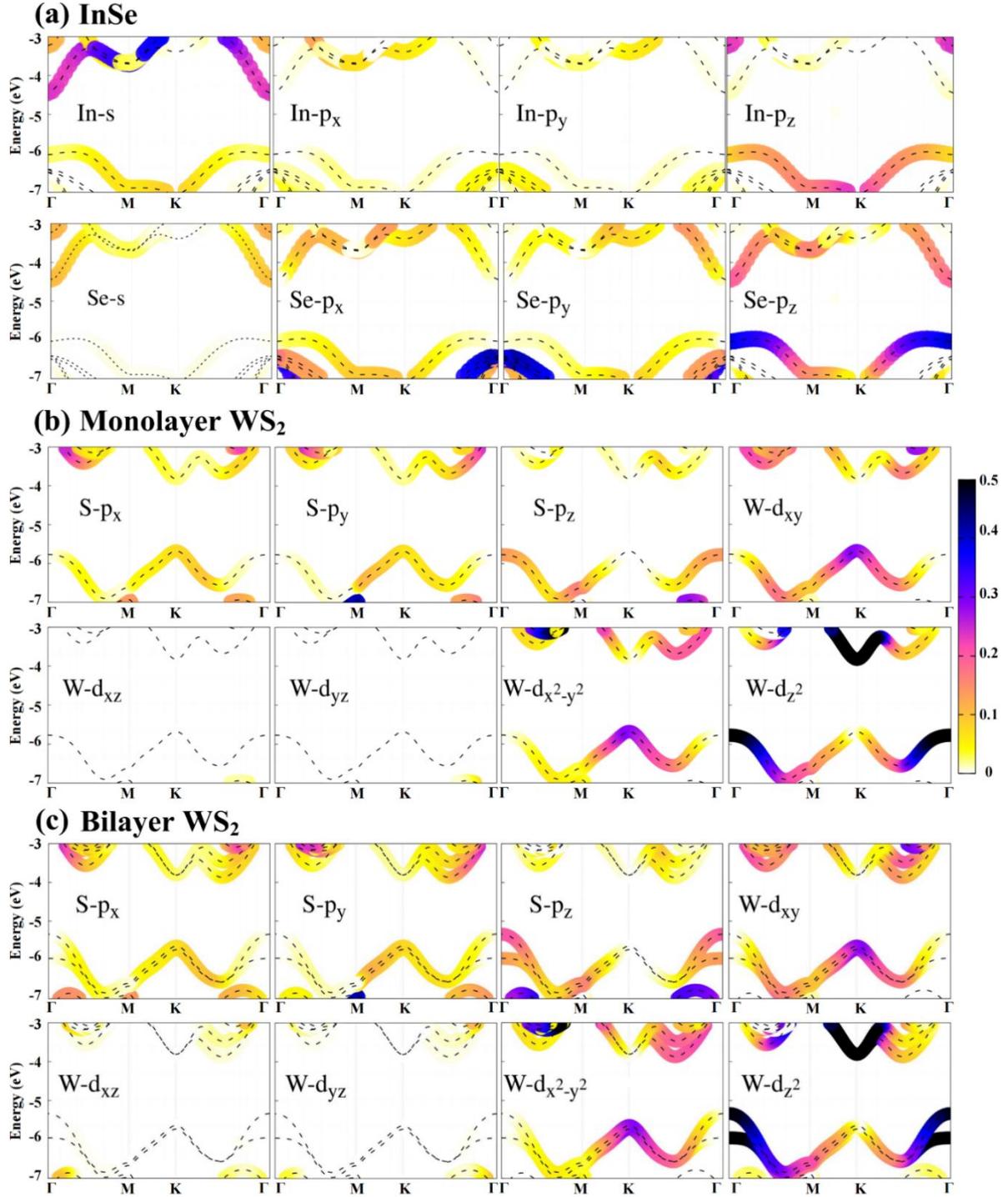

FIG. S2. Orbital-projected band structure for InSe, monolayer WS$_2$ and BL-WS$_2$. The projections are made on the sulfur *p*-orbitals and the tungsten *d*-orbitals. The scale indicates the relative weight of an orbital. The dashed lines indicate the band structure. The vacuum level was set to zero.

From monolayer (1L) $WS_2$ to BL-$WS_2$, we use the following formula to describe the influence of interlayer QB on band edge evolution: $\Delta QB = VB_\Gamma(BL) - VB_K(1L)$, where $VB_\Gamma$ is the energy of VB at $\Gamma$ point and $VB_K$ is the energy of VB at K point. The $\Delta QB$ is 0.16 eV from monolayer $WS_2$ to BL-$WS_2$. As shown in Fig. S2, the VB edge at K is composed of tungsten $d_{xy}$ and $d_{x^2-y^2}$ orbitals and sulfur $p_x$ and $p_y$ orbitals. The VB edge at $\Gamma$ is composed of sulfur $p_z$ and tungsten $d_{z^2}$ orbitals. Due to the QB interaction is sensitive to the out-of-plane orbitals, the $p_z$ orbital at the $\Gamma$-point interact strongly. As a result, the VBM of BL-$WS_2$ moves form K point in monolayer to $\Gamma$ point in bilayer.

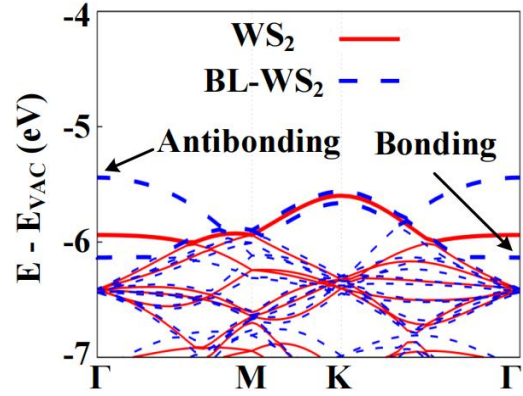

FIG. S3. Valance band structure evolution from monolayer to bilayer $WS_2$. The vacuum level was set to zero.

TABLE SII. The $k$-resolved COHP projected to the interface S-$p_z$---Se-$p_z$ pairs ($p$COHP).

| Distance of S---Se (Å) | $p$COHP of VB@$\Gamma$ | $p$COHP of VB-1@$\Gamma$ |
|---|---|---|
| 3.40 | 0.009 | 0.007 |
| 3.42 | 0.009 | 0.006 |
| 3.44 | 0.009 | 0.005 |
| 3.67 | 0.005 | 0.003 |
| 3.73 | 0.004 | 0.003 |
| 3.77 | 0.004 | 0.002 |
| 3.97 | 0.003 | 0.002 |
| **Sum of $p$COHP** | **0.043** | **0.028** |

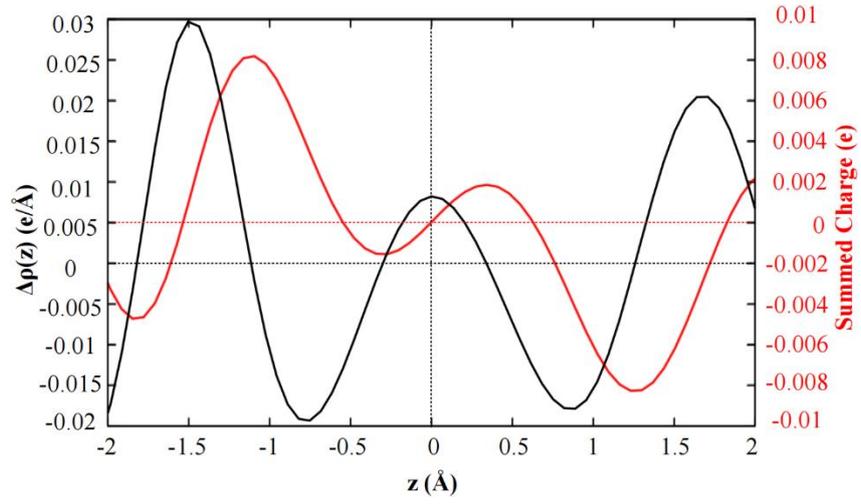

FIG. S4. The electron density difference (black line) and the summed charge (red line) along the z-direction normal to the interface of the InSe/BL-WS$_2$ vdWH.

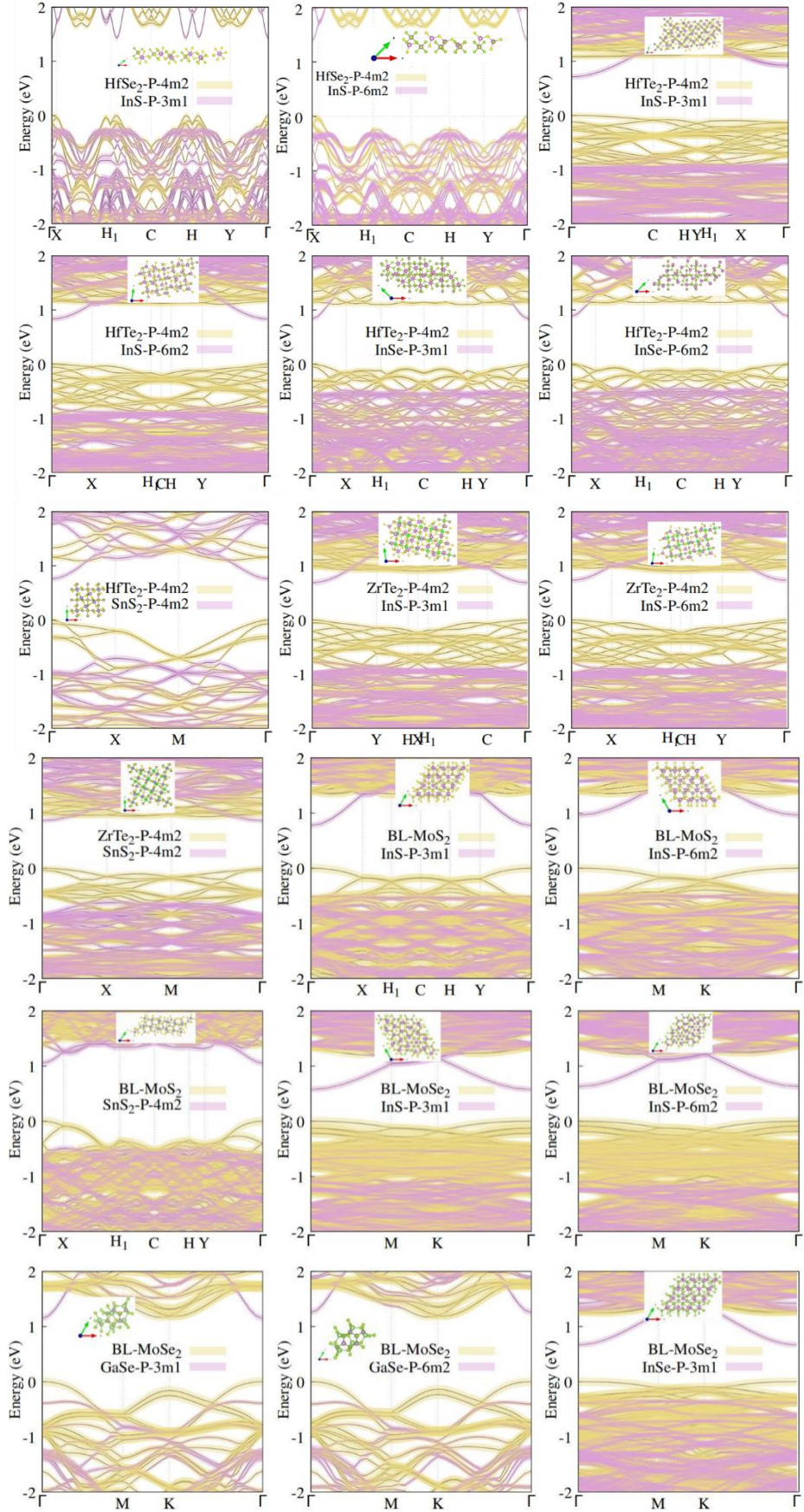

FIG. S5 (Part 1). Band structures of the 1st - 18th of the 26 robust momentum-matched type II vdWHs.

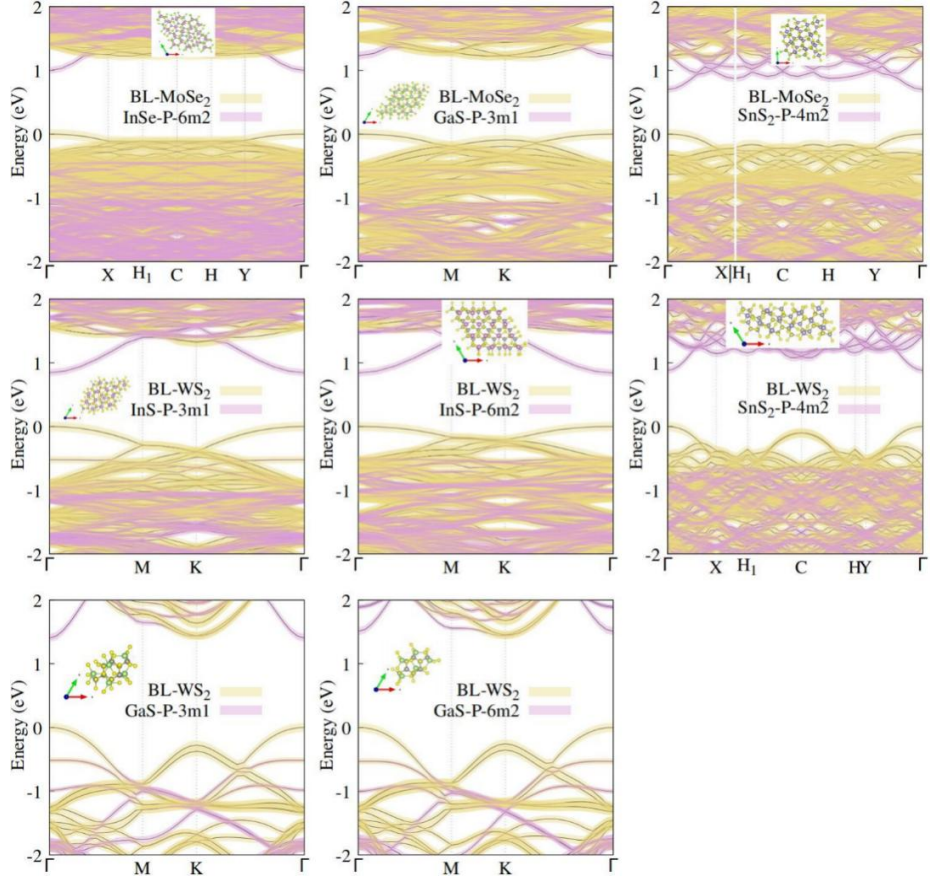

FIG. S5 (Part 2). Band structures of the 19th - 26th of the 26 screened robust momentum-matched type II vdWHs.

TABLE SIII. The VBM and CBM values with respective to the vacuum energy level, the band gaps, and the lattice constants of the 37 2DSs for the screening [2,3].

| Space group | Formula | VBM (eV) | CBM (eV) | $E_g$ (eV) | a (Å) |
|---|---|---|---|---|---|
| P-3m1 | InTe | -5.397 | -4.131 | 1.266 | 4.381 |
| P-4m2 | $TiSe_2$ | -5.977 | -5.020 | 0.957 | 3.829 |
| P-4m2 | $HfSe_2$ | -6.124 | -4.337 | 1.787 | 4.01 |
| P-3m1 | $HfSe_2$ | -5.578 | -4.940 | 0.637 | 3.773 |
| P-6m2 | $TiSe_2$ | -6.177 | -5.572 | 0.604 | 3.495 |
| P-4m2 | $ZrS_2$ | -6.687 | -4.736 | 1.951 | 3.915 |
| P-3m1 | GeSe | -5.834 | -3.541 | 2.293 | 3.666 |
| P-3m1 | $ZrS_2$ | -6.469 | -5.266 | 1.203 | 3.682 |
| P-6m2 | $TiTe_2$ | -5.303 | -5.140 | 0.162 | 3.737 |
| P-4m2 | TiTe$_2$ | -5.329 | -4.764 | 0.564 | 4.123 |
| P-4m2 | HfTe$_2$ | -5.458 | -4.256 | 1.203 | 4.282 |
| P-6m2 | HfTe$_2$ | -5.483 | -5.128 | 0.355 | 3.909 |
| P-4m2 | ZrSe$_2$ | -6.112 | -4.581 | 1.532 | 4.038 |
| P-3m1 | ZrSe$_2$ | -5.623 | -5.121 | 0.502 | 3.795 |
| P-3m1 | SnSe$_2$ | -6.051 | -5.249 | 0.802 | 3.865 |
| P-4m2 | ZrTe$_2$ | -5.449 | -4.462 | 0.986 | 4.307 |
| P-4m2 | TiS$_2$ | -6.580 | -5.294 | 1.285 | 3.676 |
| P-6m2 | ZrTe$_2$ | -5.614 | -5.162 | 0.452 | 3.923 |
| P-4m2 | HfS$_2$ | -6.708 | -4.492 | 2.216 | 3.879 |
| P-3m1 | HfS$_2$ | -6.421 | -5.106 | 1.316 | 3.652 |
| P-3m1 | GaS | -6.232 | -4.052 | 2.181 | 3.656 |
| C2/m | InSe | -5.436 | -3.772 | 1.664 | 4.099 |
| P-6m2 | GaS | -6.269 | -3.946 | 2.323 | 3.645 |
| P-6m2 | GaSe | -5.821 | -4.034 | 1.787 | 3.824 |
| P-6m2 | InS | -6.350 | -4.672 | 1.678 | 3.915 |
| P-3m1 | InS | -6.327 | -4.722 | 1.605 | 3.919 |
| P-3m1 | GaSe | -5.792 | -4.135 | 1.657 | 3.835 |
| P-4m2 | SnS$_2$ | -6.118 | -4.680 | 1.438 | 3.782 |
| P-3m1 | InSe | -5.915 | -4.575 | 1.340 | 4.085 |
| P-6m2 | InSe | -5.941 | -4.511 | 1.431 | 4.074 |
| P-4m2 | SnSe$_2$ | -5.370 | -4.511 | 0.859 | 3.953 |
| P-6m2 | InTe | -5.417 | -4.052 | 1.365 | 4.370 |
| P-6m2 | BL-MoS$_2$ | -5.521 | -4.174 | 1.347 | 3.164 |
| P-6m2 | BL-MoSe$_2$ | -5.128 | -3.929 | 1.200 | 3.295 |
| P-6m2 | BL-MoTe$_2$ | -4.811 | -3.828 | 0.983 | 3.519 |
| P-6m2 | BL-WS$_2$ | -5.333 | -3.871 | 1.462 | 3.173 |
| P-6m2 | BL-WSe$_2$ | -4.999 | -3.648 | 1.351 | 3.293 |

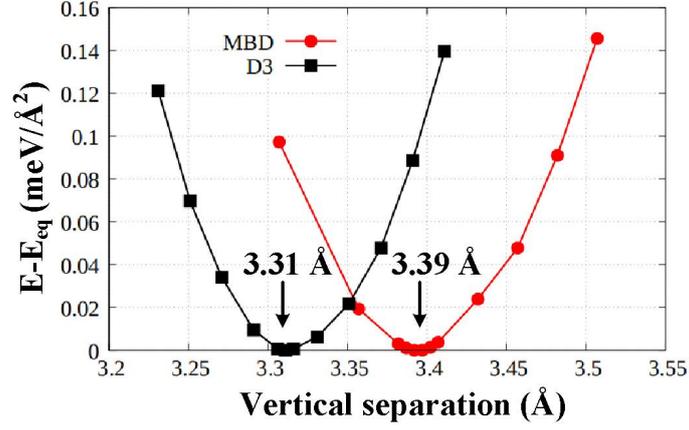

FIG S6. Energy curves, which take the energy of equilibrium interlayer distance ($E_{eq}$) as a reference, as a function of the interface vertical separation between InSe and $WS_2$ ($d_{S-Se}$) calculated with PBE+D3 and PBE+MBD methods.

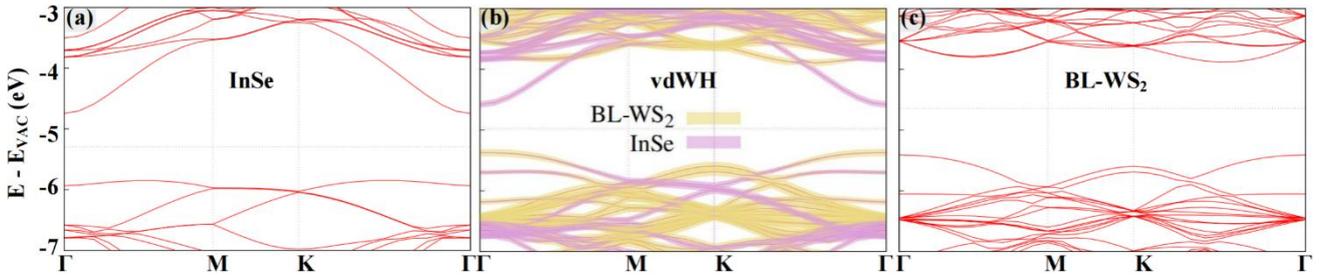

FIG S7. Band structures and alignment of the band edges (relative to the vacuum level) calculated by the PBE+MBD method. The energy reference is the vacuum levels ($E_{VAC}$) of the isolated InSe, the InSe/BL-$WS_2$ vdWH, and the isolated BL-$WS_2$, respectively. In vdWH, there are two vacuum levels on the InSe and BL-$WS_2$ sides and the vacuum level on the BL-$WS_2$ side is set to zero.

It is reported that the introduced strain of less than 5% has little impact on the quantitative estimation of QB and interface dipole [1]. In addition, as shown in Fig. S8, the type II alignment maintained taking into account of the effects of strain or not.

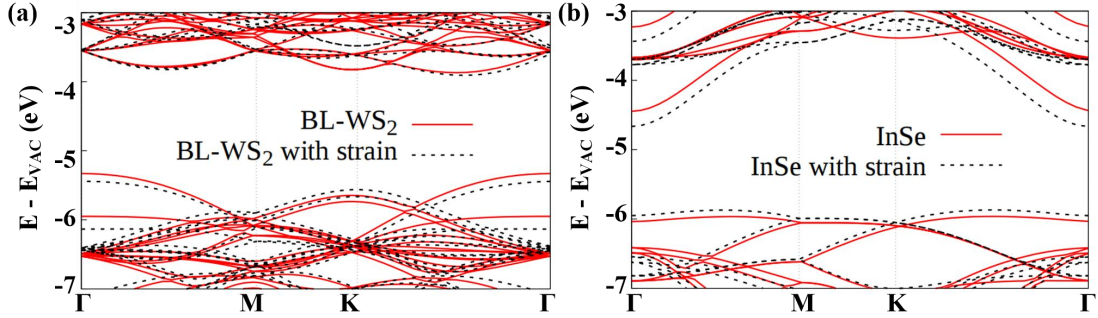

FIG S8. Electronic band structures of (a) BL-WS$_2$ without and with 1.69% strain, (b) InSe without and with 1.69% strain.

Another method to reduce the strain is to construct large supercells to obtain smaller lattice mismatches. As shown in Fig. S9, supercells of InSe-($\sqrt{19} \times \sqrt{19}$) and BL-WS$_2$-($\sqrt{31} \times \sqrt{31}$) are used to construct the vdWH, with a strain of 0.13% for the InSe/BL-WS$_2$ heterostructure.

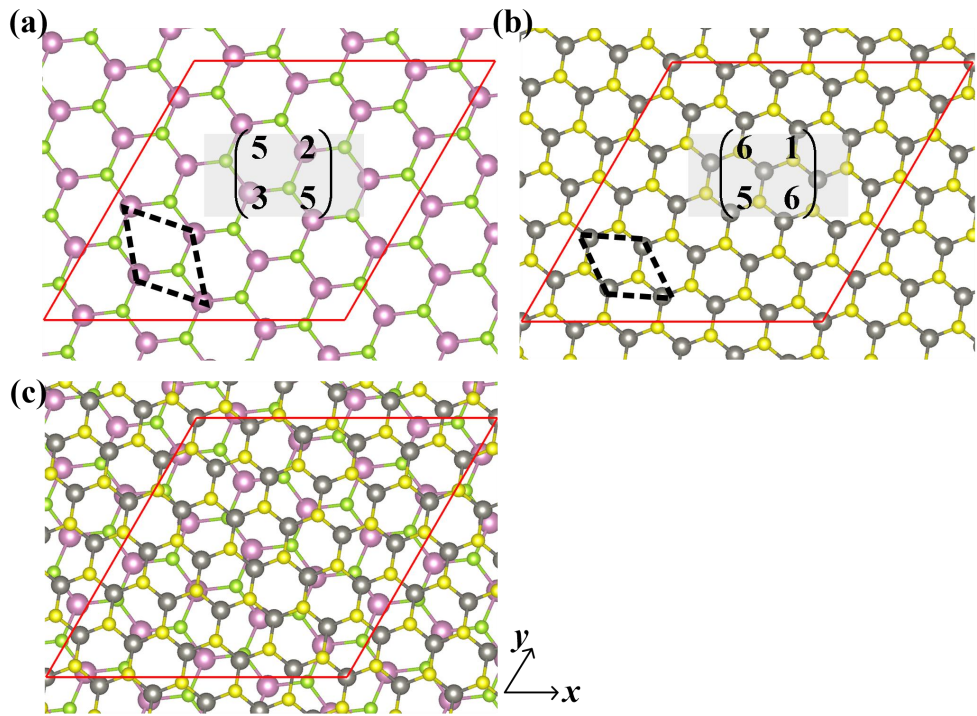

FIG S9. Geometric structures of (a) InSe with unticell (black dashed line) and supercell (red line), (b) BL-WS$_2$ with unitcell and supercell, and (c) the InSe/BL-WS$_2$ vdWH.

In the manuscript, as a balance of computational cost and accuracy, supercells of InSe-(2×2) and BL-WS$_2$-($\sqrt{7} \times \sqrt{7}$)R19.1° were used to construct the vdWH for InSe/BL-WS$_2$ heterostructure. To focus on the interfacial effects of QB and interface dipole, the same lattice constant is used for the calculations of subsystems with supercells [InSe-(2 × 2) and BL-WS$_2$- ($\sqrt{7} \times \sqrt{7}$ )R19.1 °] and vdWH. Specifically, the averaged lattice-constant of subsystem supercells is adopted. The projected band structures of isolated InSe and BL-WS$_2$ in Fig. 2c have already included the strain effect.